\documentclass[twoside,8pt]{article}
\usepackage{epsfig}
\usepackage{amsmath,amsfonts}

\def\d{\textrm{d}}
\def\e{\textrm{e}}

\newcommand{\be}{\begin{equation}}
\newcommand{\ee}{\end{equation}}
\newcommand{\bea}{\begin{eqnarray}}
\newcommand{\eea}{\end{eqnarray}}

\begin{document}

\title{Statistical measure of complexity of hard-sphere gas: applications to nuclear matter}

\author{
Ch.C. Moustakidis$^{1}$, K.Ch. Chatzisavvas$^{1,2}$, N.S. Nikolaidis$^{1,3}$, and C.P. Panos$^{1}$ \\
\\
$^{1}$Department of Theoretical Physics, Aristotle University of
Thessaloniki, \\ 54124 Thessaloniki, Greece \\
$^{2}$Informatics and Telecommunications Engineering Department,
\\ University of Western Macedonia, 50100 Kozani, Greece\\
$^{3}$Department of Automation, Faculty of Applied Technology \\
Alexander Technological Educational Institute (ATEI) of Thessaloniki \\
57400 Thessaloniki, Greece\\
}

\maketitle

\begin{abstract}
We apply  the statistical measure of complexity, introduced by
L\'{o}pez-Ruiz, Mancini and Calbet to a  hard-sphere dilute Fermi
gas whose particles interact via a repulsive hard-core potential.
We employ the momentum distribution of this system to calculate
the information entropy, the disequilibrium and the statistical
complexity. We examine possible connections between the particle
correlations and energy of the system with those information and
complexity measures. The hard-sphere model serves as a test bed
for concepts about complexity.

\vspace{0.3cm}

PACS number(s): 05.30.Fk, 89.70.Cf, 05.30.-d, 05.90.+m  \\

Keywards: Momentum Distribution, Fermi Systems, Nuclear Matter,
Hard-Sphere Gas, Information Entropy, Statistical Complexity,
Correlations.

\end{abstract}

%%%%%%%%%%%%%%%%%%%%%%%
\section{Introduction}
%%%%%%%%%%%%%%%%%%%%%%%
There has been a remarkable growth in research focused on
complexity and in general on information theories in recent years
in a variety of fields \cite{Pines-99} including physics
\cite{Bialynicki75}, chemistry \cite{Seurs-80}, biology
\cite{Adami-04}, neuroscience \cite{Borst-99}, mathematics
\cite{Cover} and computer science \cite{Nillsen}.   In particular
there are various applications in quantum many-body systems e.g.
atoms \cite{Chatzisavvas05,Chatzisavvas07a,Chatzisavvas07b},
nuclei \cite{Panos97,Massen98,Moustakidis05,Chatzisavvas-05},
atomic clusters \cite{Massen98}, bosonic traps \cite{Massen98}
e.t.c. In fact, they lead to the clarification of basic quantum
concepts and provide results about the information content of
systems according to various definitions e.g. Shannon
 information entropy \cite{Shannon48}, Fisher \cite{Fisher25} information, Onicescu
information energy \cite{Onicescu66} e.t.c. Also they represent a
suitable framework, via a probabilistic description, to assess the
presence of interactions, correlations with experimentally
measurable quantities, the derivation of universal relations e.t.c
\cite{Panos97,Massen98,Moustakidis05,Shannon48,Fisher25,Onicescu66,Kullback59,Renyi61,
  Gadre84,Ohya93,BAB95,Nagy96,Majer96,
Ghosh84,Garba06,Frieden04,Patil07,Liu07,Luzanov07,Antolin08,Sagar08}.
Thus, traditional methods can be extended by an alternative
information-theoretic way and give new insights for the treatment
of simple quantum systems as well as more complicated many-body
ones. A recent and important advance is to calculate several
complexity measures, based on a probabilistic description via
previous experience on information entropy in order to quantify
statistical indicators of complex behavior in different systems
scattered in a broad spectrum of fields
\cite{Chatzisavvas05,Chatzisavvas07a,Chatzisavvas07b,
  Lopez95,Catalan02,Sanudo09,Landsberg-98,Shiner-99,Crutch98,Crutch00,Binder00,Martin03,
Panos-09,Lopez05,Yamano04,Plastino96,Borgoo07,Montogomery08,Calbet-07,
Angulo08,Angulo08b,Sanudo08,Sanudo08b,Chatzi-09,Roza-09-a,Ruiz-09,Antolin09}.
Related research started connecting the above measures with
experiment e.g. Fisher information entropy has been found to
correlate with  the ionization potential and dipole polarizability
in atoms \cite{Chatzisavvas07a} and also complexity in a
correlated Fermi gas has been connected with the specific heat
$C_V$ e.t.c. \cite{Moustakidis-010}.

The statistical measure of complexity $C_{LMC}$ introduced by
L\'{o}pez-Ruiz, Calbet and Mancini (LMC) \cite{Lopez95} is defined
in the form of the product $C_{\rm LMC}=S D$. Here, $S$ is the
information entropy i.e. the information content of the system,
while $D$ is the disequilibrium i.e. the distance to the
equilibrium probability distribution. Although complexity is a
multi-faced quantity and several definitions of complexity
measures have been proposed, the LMC  measure has been employed
recently in various studies for the following reasons: it exhibits
the correct asymptotic properties of a well-behaved measure of
complexity, as expected by intuition e.g. it vanishes for the two
extreme cases of a perfect crystal (complete order) and ideal gas
(complete disorder). In addition it is easily calculable for a
quantum system, described by its very nature by probability
densities leading to a feasible calculation of its basic factors
$S$ and $D$. Other definitions of complexity, although sometimes
may be considered that they describe certain aspects of complex
systems in a satisfactory way, they have other disadvantages e.g.
Kolmogorov's algorithmic complexity is hard to compute. It is
defined as the length of the shortest (optimum) program needed to
describe the system, a goal difficult to attain and prove
\cite{Kolmogorov-65}. Our approach is a pragmatic one: we start
from the LMC definition, which is relatively easily calculable and
hope to improve in the future, by a assessing the obtained
results, comparing with other definitions of complexity e.g. the
SDL measure according to Shiner, Davison and Landsber
\cite{Shiner-99}.

The initial definition of $C_{LMC}$ has been slightly modified in
a suitable way by Catalan \emph{et al} \cite{Catalan02}, leading
to the form $C=\e^S D$ applicable to systems described by either
discrete or continuous probability distributions. In
\cite{Catalan02} it was shown that the results in both, discrete
and continuous cases, are consistent: extreme values of $C$ are
observed for distributions characterized by a peak superimposed
onto a uniform sea. Moreover, $C$ should be minimal, when the
system reaches equipartition and the minimum value of $C$ is
attained for rectangular (uniform) density probabilities giving
the value $C=1$. Additionally, $C$ is not an upper bounded
function and can become infinitely large.

The motivation of the present work, is to extend our previous
study on complexity measures of uniform Fermi systems
\cite{Moustakidis05}, by employing  the complexity measure
proposed by L\'{o}pez-Ruiz et al. \cite{Lopez95}, using
probability distributions in momentum space. In uniform systems
the density $\rho=N/V$ is a constant and the interaction of the
particles is reflected to the momentum distribution which deviates
from the $theta$ function form of the ideal Fermi-gas model. Our
aim is to connect $C$, a measure based on a probabilistic
description and the shape of the corresponding momentum
distributions to the phenomenological parameters introducing the
inter-particle correlations.

The study of uniform quantum systems (both fermionic or bosonic)
in momentum space  is very important. Very interesting phenomena
such as superfluidity, superconductivity, Bose-Einstein
condensation e.t.c are observed and also well defined in momentum
space. Thus, it is interesting  to concentrate our study on the
connection between complexity, defined in momentum space to the
correlated behavior of a fermionic (or bosonic ) system, by
employing the simple, but  effective, hard-sphere model. Our
specific application is nuclear matter. The basic model with
hard-spheres is a suitable starting point in order to assess the
relevance of various concepts and definitions of complexity. The
present application in a correlated Fermi system like nuclear
matter is facilitated by our previous experience. We use the
simplest potential of a hard-sphere interaction with a hard core
radius. The outline of the present work is the following: In
Sec.~2 we present the momentum distribution and information
measures employed to quantify complexity, while Sec.~3 contains
our numerical results and discussion. Finally, in Sec.~4 we
exhibit our conclusions.

%%%%%%%%%%%%%%%%%%%%%%%%%%%%%%%%%%%%%%%%
\section{Momentum distribution and information measures}
%%%%%%%%%%%%%%%%%%%%%%%%%%%%%%%%%%%%%%%%
We adopt the formalism employed in our previous work
\cite{Moustakidis05,Moustakidis-010}, adjusted here to a
hard-sphere gas and specializing in nuclear matter.
\subsection{Momentum distribution}
%%%%%%%%%%%%%%%%%%%%%%%%%%%%%%%%%%
The momentum distribution (MD) of an interacting Fermi system is
given in general  by the relation
\begin{equation}
n(k)=\frac{1}{V_k}\left\{
\begin{array}{rr}
n_{<}(k), \quad  \text{for $k>k_F$}\\
n_{>}(k), \quad \text{for $k<k_F$}
\end{array} \right.
\label{nk-1}
\end{equation}
where $V_k=\frac{4}{3}\pi k_F^3$. The Fermi wave number $k_F$ is
related with the constant density $\rho=N \rho_0=3/(4\pi r_0^3)$
as follows
\begin{equation}
    k_F=\left( \frac{6 \pi^2 \rho}{\nu} \right)^{1/3}= \
    \left(\frac{9 \pi}{2 \nu}\frac{1}{r_0^3} \right)^{1/3}. \label{kfermi}
\end{equation}
The normalization of $n(k)$ obeys the relation
\begin{equation}
\left(\frac{4\pi}{3}k_F^3\right)^{-1} \int n(k) d{\bf k}=1.
\label{norm-1}
\end{equation}
The simplest form for $n(k)$ appears for an ideal Fermi gas. In
this case $n(k)$ is just a step function
\begin{equation}
    n_0(k)=\frac{1}{V_k}\, \theta(k_F-k). \label{MD-IG}
\end{equation}
The potential of the hard-sphere interaction is defined as
follows,
\begin{equation}
V(r)=\left\{
\begin{array}{rr}
\infty, \quad  \text{for $r<c$}\\
0, \quad \text{for $r>c$},
\end{array} \right.
\label{hsi}
\end{equation}
where $c$ denotes the hard core radius.

The momentum distribution of a hard-sphere dilute Fermi gas had
previously been calculated by Czyz and Gottfried \cite{Czyz-61}
and also by Sartor and Mahaux \cite{Sartor-80}. The above authors
have studied a low density Fermi gas, whose particles interact via
a repulsive hard core potential of the form (\ref{hsi}). In this
model, the quantities of interest can be expanded in powers of the
parameter $(k_Fc)$.

The analytical expressions for the dimensionless $n(k)$ in the
hard-sphere Fermi gas model have the following form for  $k<k_F$
\cite{Sartor-80}
\begin{eqnarray}
n_{<}(x)=&1&-\frac{\nu-1}{3\pi^2 x}(k_Fc)^2 \left[ \frac{}{}(7
\ln2-8)x^3+(10-3\ln 2)x \right. \nonumber \\
&+&\left.  2\ln \frac{1+x}{1-x}-2(2-x^2)^{3/2} \ln
\frac{(2-x^2)^{1/2}+x}{(2-x^2)^{1/2}-x} \right ], \label{nk-l}
\end{eqnarray}
where $x=k/k_F$ and $\nu=4$. For $1<x<\sqrt{2}$:
\begin{eqnarray}
n_{>}(x)&=&\frac{\nu-1}{6\pi^2 x}(k_Fc)^2\left((7x^3-3x-6)\ln
\frac{x-1}{x+1}+(7x^3-3x+2)\ln2 -8x^3+22x^2+6x
\right. \\
&-&\left. 24+2(2-x^2)^{3/2}\left[
\ln\frac{2+x+(2-x^2)^{1/2}}{2+x-(2-x^2)^{1/2}}+
\ln\frac{1+(2-x^2)^{1/2}}{1-(2-x^2)^{1/2}}-
2\ln\frac{x+(2-x^2)^{1/2}}{x-(2-x^2)^{1/2}}\right]\right).
\nonumber \label{nk-r-1}
\end{eqnarray}
For $\sqrt{2}<x<3$:
\begin{eqnarray}
n_{>}(x)&=&\frac{\nu-1}{6\pi^2 x}(k_Fc)^2\left((7x^3-3x-6)\ln
\frac{x-1}{x+1}+(7x^3-3x+2)\ln2 -8x^3+22x^2+6x
\right. \\
&-&\left. 4(x^2-2)^{3/2}\left[ \tan^{-1}\frac{x+2}{(x^2-2)^{1/2}}+
\tan^{-1}\frac{1}{(x^2-2)^{1/2}}-
2\tan^{-1}\frac{x}{(x^2-2)^{1/2}}\right]\right). \nonumber
\label{nk-r-2}
\end{eqnarray}
For $3<x$:
\begin{eqnarray}
n_{>}(x)&=&2\frac{\nu-1}{3\pi^2 x}(k_Fc)^2\left(2\ln
\frac{x+1}{x-1}-2x+(x^2-2)^{3/2}
\right. \\
&\times&\left. \left[2\tan^{-1}\frac{x}{(x^2-2)^{1/2}}-
\tan^{-1}\frac{x-2}{(x^2-2)^{1/2}}-\tan^{-1}\frac{x+2}{(x^2-2)^{1/2}}
\right ]\right). \nonumber \label{nk-r-3}
\end{eqnarray}

Another characteristic quantity, used as a measure of the strength
of correlations of uniform Fermi systems, is the discontinuity,
$Z$, of the momentum distribution at $k/k_F=1$. It is defined as
\begin{equation}
    Z=n(1^-)-n(1^+).
    \label{Z-1}
\end{equation}

The behavior of the momentum distribution, as a function of
$x=k/k_F$ for various values of the correlation parameter $k_Fc$
is shown in Fig.~1. The discontinuity $Z$ is also displayed in
each case. For ideal Fermi systems $Z=1$, while for interacting
ones $Z<1$. In the limit of very strong interaction where $Z
\rightarrow 0$, there is no discontinuity in the momentum
distribution of the system. The quantity ($1-Z$) measures the
ability of correlations to deplete the Fermi sea by exciting
particles from states below it (hole states) to states above it
(particle states) \cite{Flyn84}.

The asymptotic behavior of $n(1^-)$ for $x->1^{-}$ reads
\cite{Sartor-80}
\begin{equation}
n(1^-)\approx 1-\frac{2}{3\pi^2}(\nu-1)(k_Fc)^2\left[3\ln
2+1-3(x-1)\ln \mid x-1\mid +(6\ln 2 -\frac{15}{2})(x-1)\right],
\label{nim-1}
\end{equation}
while for $x->1^{+}$
\begin{equation}
n(1^+)\approx \frac{2}{3\pi^2}(\nu-1)(k_Fc)^2\left[\frac{}{}3\ln
2-1-3(x-1)\ln (x-1) -(6\ln 2 -7)(x-1)\right]. \label{nip-1}
\end{equation}
It is worthwhile to note the existence of  a logarithmic
singularity in the function $n(k)$ at $k=k_F$, a general feature
of normal Fermi systems.

The discontinuity $Z$, according to Eqs~(\ref{Z-1}) and
(\ref{nim-1}), (\ref{nip-1}) is given by
\begin{equation}
Z=1-\frac{4}{\pi^2}\ln 2(\nu-1)(k_Fc)^2. \label{Z-2}
\end{equation}
The energy per particle $E$ of the ground state of
$\nu=4$-component fermion fluid of hard-spheres, in the
low-density expansion, has been derived in Refs.~\cite{Baker-82}.
Accordingly,  the energy $E$, in units of the ideal gas energy
$E_0$, is given by \cite{Baker-82}
\begin{equation}
E/E_0=e(y)\simeq 1+D_1y+D_2y^2+D_3y^3+D_4y^4\ln y,\qquad
y=k_Fc,\quad E_0=\frac{3}{5}\frac{\hbar^2 k_F^2}{2m}. \label{En-1}
\end{equation}
where the coefficients $D_i$ are given in Table VI of
Ref.~\cite{Baker-82}. It is one of the aims of the present work to
investigate the connection between the various information
measures and complexity with experimental quantities (as the
ground state energy and the discontinuity $Z$). In addition, we
intend to produce not only qualitative but mostly  quantitative
results, by connecting the strength of the correlations with the
above measures.

%%%%%%%%%%%%%%%%%%%%%%%%%%%%%%%%%%%
\subsection{Information measures}
%%%%%%%%%%%%%%%%%%%%%%%%%%%%%%%%%%%
The information entropy in momentum space is given by the relation
\begin{equation}
    S_k=-\int n(k) \ln{n(k)} \, \d{\bf k} . \label{IE-Sk-1}
\end{equation}
So, for an ideal Fermi gas, using Eq. (\ref{MD-IG}), $S_{k}$
becomes
\begin{equation}
S_k=S_0=\ln V_k=\ln \left(\frac{6 \pi^2}{\nu}\frac{1}{r_0^3}
\right). \label{IE-Sk-2}
\end{equation}
For correlated Fermi systems, $S_k$, can be found from Eq.
(\ref{IE-Sk-1}) by employing Eq.~(\ref{nk-1}). $S_k$ is written
now \cite{Moustakidis05}
\begin{equation}
S_k=\ln V_k-\frac{4 \pi}{V_k}\left( \int_{0}^{k_F^{-}} k^2
n_{<}(k)\ln n_{<}(k)  dk+ \int_{k_F^{+}}^{\infty} k^2 n_{>}(k) \ln
n_{>}(k)  dk \right). \label{Cor-Sk-1}
\end{equation}
The correlated entropy $S_k$ has the form
\begin{equation}
S_k=S_0+S_{\rm cor}, \label{Cor-S-1}
\end{equation}
where $S_0$ is the uncorrelated entropy given by
Eq.~(\ref{IE-Sk-2}) and $S_{\rm cor}$ is the contribution of the
particle correlations to the entropy. That contribution can be
found from the expression
\begin{equation}
S_{\rm cor}=-3\left( \int_{0}^{1^{-}} x^2 n_{<}(x)\ln n_{<}(x) dx+
\int_{1^{+}}^{\infty} x^2 n_{>}(x)\ln n_{>}(x) dx \right),
\label{S-cor}
\end{equation}
where $x=k/k_F$.

The disequilibrium $D_k$ (or information energy, defined by
Onicescu \cite{Onicescu66}), in momentum space   as another
functional of a probability distribution, in our case $n(k)$ is
given by the relation
\begin{equation}
D_k=\int n^2(k) d{\bf k} . \label{D-1}
\end{equation}
For an ideal Fermi gas, using Eq. (\ref{MD-IG}), becomes
\begin{equation}
D_k=D_0=\frac{1}{V_k}. \label{D-ideal}
\end{equation}
In the case of correlated Fermi systems, $D_k$ is written as
\begin{equation}
D_k=\frac{1}{V_k} \frac{4\pi}{V_k}\left( \int_{0}^{k_F^{-}} k^2
n^2_{<}(k) dk+ \int_{k_F^{+}}^{\infty} k^2 n^2_{>}(k)dk \right).
\label{D-Cor-1}
\end{equation}
The correlated disequilibrium $D_k$ is
\begin{equation}
D_k=D_0  D_{\rm cor}, \label{D-Cor-2}
\end{equation}
where $D_0$ is  given in  Eq. (\ref{D-ideal}) and $D_{cor}$  can
be found also from the expression
\begin{equation}
D_{cor}=3\left( \int_{0}^{1^{-}} x^2 {n}^2_{<}(x)dx+
\int_{1^{+}}^{\infty} x^2 {n}^2_{>}(x) dx \right). \label{D-cor-3}
\end{equation}
The statistical complexity measure, proposed by Catalan et. al.
\cite{Catalan02}, in momentum space,  is defined as
\begin{equation}
C_{\rm LMC}=C=D_k H_k, \label{C-1}
\end{equation}
where  $H$ represents the information content of the system
defined as
\begin{equation}
H_k=\e^{S_k}. \label{H-1}
\end{equation}
It is easy to show that
\begin{equation}
C=C_0  C_{\rm cor}=D_{\rm cor} \e^{S_{\rm cor}}, \qquad C_0=D_0
\e^{S_0}=1. \label{C-2}
\end{equation}
The physical meaning of Eq.~(\ref{C-2}) is clear. In the case of
an ideal Fermi gas (see Eq.~(\ref{MD-IG})) $C$ is minimal with the
value $C_0=1$ (see also \cite{Catalan02}). Moreover as pointed out
in Ref.~\cite{Catalan02}, $C$ is not an upper bounded function and
can therefore become infinitely large. From the above analysis it
is clear that complexity $C$ is an accounter of correlations in an
infinite Fermi system. Thus, the next step is to try to find the
connection between $C$ and the correlation parameters of the
systems. The correlation invoke diffusion of the momentum
distribution and we expect this effect to be reflected on the
values of $C$.

%%%%%%%%%%%%%%%%%%%%%%%%%%%%%%%%%%%%%%%%%%%
\section{Results and discussion}
%%%%%%%%%%%%%%%%%%%%%%%%%%%%%%%%%%%%%%%%%%%
The behavior of the momentum distribution, as a function of
$k/k_F$, for various values of the wound parameter $k_Fc$ is shown
in Fig.~1. The discontinuity $Z$ is also displayed in each case.
For ideal Fermi systems $Z=1$, while for interacting ones $Z<1$.
In the limit of very strong interaction $Z=0$, there is no
discontinuity on the momentum distribution of the system.

The calculated values of $S_{\rm cor}$, $D_{\rm cor}$ and $C$ for
nuclear matter versus the correlation  parameter $k_Fc$ are
displayed in Fig.~2. $S_{\rm cor}$ and $C$ increase with $y$,
while $D_{\rm cor}$ decreases. We fitted the numerical values of
the above quantities, with simple functions of $y=k_Fc$ and we
find respectively the following formulae
\begin{equation}
    S_{\rm cor}(y)=\alpha y^{\beta}, \qquad \alpha=2.16379, \quad
    \beta=1.67053. \label{Scor-fit-1}
\end{equation}
\begin{equation}
D_{\rm cor}(y)=1+\alpha y^{\beta}, \qquad \alpha=-0.79871, \quad
\beta=1.83155. \label{Dcor-fit-1}
\end{equation}
\begin{equation}
C(y)=1+\alpha y^{\beta}, \qquad \alpha=1.68358, \quad
\beta=1.67566. \label{Ccor-fit-1}
\end{equation}
The values of the parameters $\alpha$, $\beta$ and  $\gamma$, for
each case, have been selected by a least squares fit (LSF) method.
It is worthwhile to notice that for  an ideal Fermi gas there is
an upper limit for $C$ ($C_{max}\simeq 1.5845$)
\cite{Moustakidis-010}. However, for an interacting Fermi gas
there is no such constraint (at least to the region  under
consideration in the present work). It is obvious that the
calculated values of complexity reflect the different way that the
interaction or temperature affect the trend of the momentum
distribution.

The quantity ($1-Z$) measures the ability of correlations to
deplete the Fermi sea by exciting particles from states below it
(hole states) to states above it (particle states) \cite{Flyn84}.
The dependence of $S_{\rm cor}$, $D_{\rm cor}$ and  $C$ on the
quantity $(1-Z)$ is shown in Fig.~3. It is seen that $S_{\rm cor}$
and $C$ are increasing functions of $(1-Z)$, while $D_{\rm cor}$
is a decreasing one, as a direct consequence of the dependence of
the above quantities on the correlation parameter $k_Fc$. That
dependence can be reproduced very well by simple expressions as in
Eqs.~(\ref{Scor-fit-1}), (\ref{Dcor-fit-1}) and (\ref{Ccor-fit-1})
replacing  $k_Fc$ by $(1-Z)$
\begin{equation}
    S_{\rm cor}(Z)=\alpha (1-Z)^{\beta}, \qquad \alpha=2.49614, \quad
    \beta=0.83527. \label{Scor-fit-2}
\end{equation}
\begin{equation}
D_{\rm cor}(Z)=1+\alpha (1-Z)^{\beta}, \qquad \alpha=-0.93413,
\quad \beta=0.91574. \label{Dcor-fit-2}
\end{equation}
\begin{equation}
C(Z)=1+\alpha (1-Z)^{\beta}, \qquad \alpha=1.94305, \quad
\beta=0.83784. \label{Ccor-fit-2}
\end{equation}

From the above analysis we can conclude that LMC complexity $C$
can be employed as a measure of the strength of correlations in
the same way the wound and the discontinuity parameters are used.
An explanation of the above behavior of $C$ is the following: The
effect of nucleon correlations is the departure from the step
function form of the momentum distribution (ideal Fermi gas) to
the one with long tail behavior for $k>k_F$. The diffusion of the
distribution leads to a decrease of the order of the system (the
disequilibrium $D_k$ decreases and the information entropy $S_k$
increases accordingly). In total, the contribution of $S_k$ in $C$
dominates over the contribution of $D_k$ and thus the complexity
increases with the correlations (at least in the region under
consideration).

It is one of the aims of the present work  to connect $C$, a
measure  of complexity based on a probabilistic description and
the shape of the corresponding momentum distributions  with other
quantities like the ground state energy per particle $E$. In view
of the above, we display in Fig.~3 the dependence of $S_{cor}$,
$D_{cor}$ and $C$, as well as the energy fraction $e(y)$, on the
correlation parameter $k_Fc$. The dependence of $S_{cor}$,
$D_{cor}$ and $C$ on $e(y)$, as displayed in Fig.~4, is in a very
good approximation linear. The fitted expressions are the
following:
\begin{equation}
{\rm e}=1.0479+1.27353 S_{cor}, \label{s-e}
\end{equation}
\begin{equation}
{\rm e}=4.6861-3.5985 D_{cor}, \label{d-e}
\end{equation}
\begin{equation}
{\rm e}=-0.5847+1.6358 C. \label{c-e}
\end{equation}
In total we observe an empirical connection of the energy with
$S_{cor}$, $D_{cor}$ and $C$ calculated employing information
entropy, which, by definition, is not related directly to the
energy of the system, in contrast to the traditional concept of
thermodynamic entropy. The above results confirm our recent
finding, according to similar lines, that there is also a
connection between the "energy-like" quantity specific heat $C_V$
of an ideal electron gas with the complexity $C$
\cite{Moustakidis-010}.

Finally, in Fig.~5 we compare the present results with those taken
from the Low Order Approximation (LOA) method. Thus the momentum
distribution of nuclear matter is evaluated by employing the LOA
and the MD takes the form \cite{Flyn84}
\begin{equation}
n_{\rm LOA}(k)=\theta(k_F-k) \left[ 1-k_{\rm dir}+Y(k,8) \right]+
8 \left[ k_{\rm dir}Y(k,2)-[Y(k,4)]^2 \right], \label{mn-mom-1}
\end{equation}
where
\begin{equation}
    c_{\mu}^{-1}Y(k,\mu)=
    \frac{\e^{-\tilde{k}_{+}^{2}}-\e^{-\tilde{k}_{-}^{2}}}{2\tilde{k}}
    +\int_{0}^{\tilde{k}_{+}} \e^{-y^2} \, \d y +
    {\rm sgn}(\tilde{k}_{-}) \int_{0}^{\mid \tilde{k}_{-} \mid} \e^{-y^2} \d y,
%\label{}
\end{equation}
and
\begin{equation}
    c_{\mu}=\frac{1}{8\sqrt{\pi}}\left(\frac{\mu}{2}\right)^{3/2},
    \quad \tilde{k}=\frac{k}{\beta \sqrt{\mu}} , \quad
    \tilde{k}_{\pm}=\frac{k_{F}\pm k}{\beta \sqrt{\mu}}, \quad
    \mu=2,4,8.
%\label{}
\end{equation}
while ${\rm sgn}(x)=x/\mid x \mid$. The dimensionless  wound
parameter $k_{\rm dir}$ can serve as a rough measure of
correlations and the rate of convergence of the cluster expansion
is defined as
\begin{equation}
    k_{\rm dir}=\rho \int [f(r)-1]^2 \, \d {\bf r}. \label{eq-kdir}
\end{equation}
The normalization condition for the momentum distribution is
\begin{equation}
\int_{0}^{\infty} n_{\rm LOA}(k)k^2 \d k=\frac{1}{3}\, k_{F}^{3}.
\end{equation}
The following relation between the wound parameter $k_{\rm dir}$
and the correlation parameter $\beta$
\begin{equation}
k_{\rm dir}=\frac{1}{3 \sqrt{2\pi}}\left(\frac{k_F}{\beta}
\right)^3.
%\label{}
\end{equation}
It is clear that large values of  $k_{\rm dir}$ imply strong
correlations and simultaneously poor convergence of the  cluster
expansion. In the numerical calculations the correlation parameter
$\beta$ is in the interval: $1.01 \leq \beta \leq  2.482 $. That
range corresponds to $ 0.3 \geq k_{\rm dir}  \geq 0.02 $ and this
is  reasonable, in the case of nuclear matter \cite{Flyn84}.
However, the origin of the two methods is different and as a
consequence they influence in a different way the momentum
distribution. Nevertheless,  we found that $S_{cor}$, $D_{cor}$
and $C$ exhibit a similar trend as a functions of the parameter
$1-Z$.

\section{Conlusions}
In conclusion, we calculate information and complexity measures of a uniform
Fermi system, like nuclear matter, in the framework of the
hard-sphere model. The effect of correlations is connected
intuitively with the concept of complexity, in a qualitative, and
somehow vague way as stated in \cite{Moustakidis-010} as well. In
fact, it turns out that all information measures used by us, show
a strong dependence on the correlation parameter $k_Fc$ as well as
on the Fermi discontinuity $Z$. The most distinctive feature is
the occurrence of an almost linear dependence between information
measures and complexity  on energy. The above statement is in
keeping with the recent finding of the existence of an empirical
connection between the specific heat and complexity
\cite{Moustakidis-010}.  However, the applicability of our
approach is much wider than nuclear matter, since the impenetrable
hard spheres (not overlapping in space) can simulate the extremely
strong repulsion that atoms and spherical molecules  feel at very
small distances. Thus, the significance of a suitable
quantification of complexity emerges in statistical mechanics of
fluids and  solids. We do not claim that our approach is the only
or more important one, but so far our results are interesting and
encouraging. We stress again that the proposed LMC measure of
complexity is by its definition an appropriate one and
specifically tailored for systems described statistically, through
a probability distribution. Since "information is physical"
\cite{DiVincenzo}, it is promising to examine how far this
quotation goes, in the sense that calculations employing a good
measure of information content of a quantum system (and
consequently of complexity) are expected to give, at least,
interesting results of physical relevance. Landauer dedicated his
research on similar ideas. Hence, in our present work and previous
ones, we proceed towards a numerical quantification of complexity.
One of our goals is to examine, as a first step, whether a
particular definition of complexity is reasonable and robust
enough e.g. if one increases the value of one parameter (or
parameters) describing a quantum system, the corresponding value
of complexity increases accordingly. This would satisfy minimally
intuition about complexity and validate its definition.

%%%%%%%%%%%%%%%%%%%%%%%%%%%%%%%%%%%%%%%%%%%%%%%%%%%%%%%%%%%%%%%%%%%%%%%%%%
%%%%%%%%%%%%%%%%%%%%%%%%%%%%%%%%%%%%%%%%%%%%%%%%%%%%%%%%%%%%%%%%%%%%%%%%%%

%FIGURES
%%%%%%%%%%%%%%%%%%%%%%%%%%%%%%%%%%%%%%%%%%%%%%%%%%%%%%%%%%%%%%%%%%%%%%
%FIGURE-1
\begin{figure}
\centering
\includegraphics[height=7.0cm,width=7.0cm]{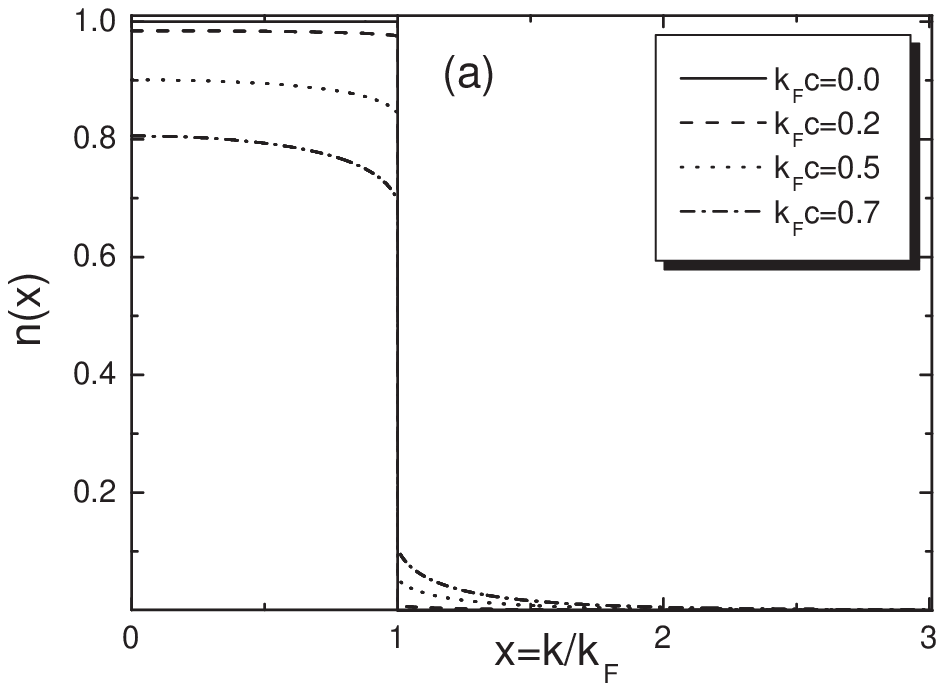}\
\includegraphics[height=7.0cm,width=7.0cm]{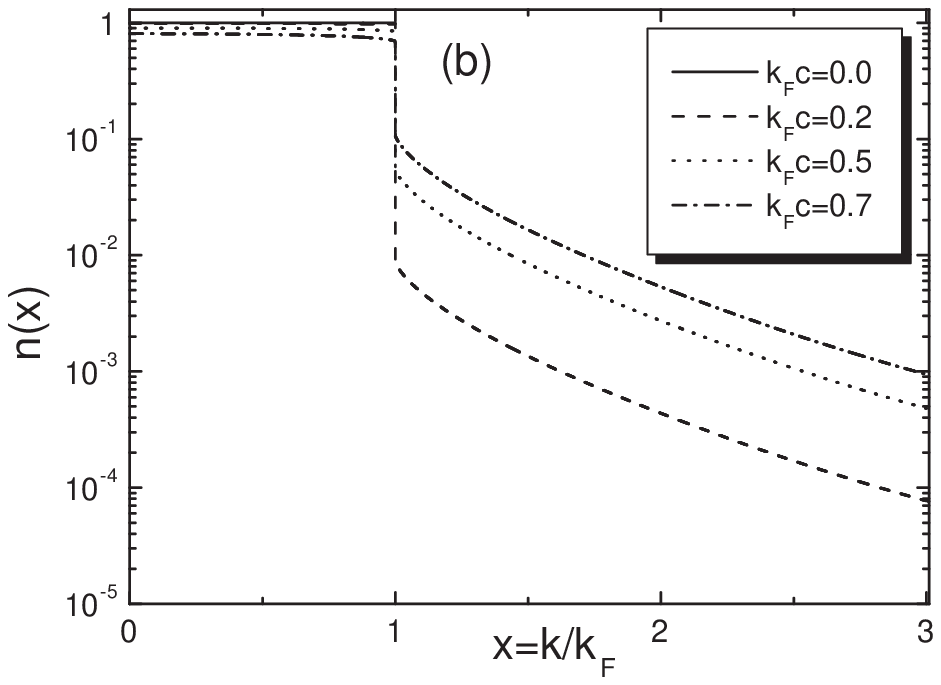}\
\caption{(a) The momentum distribution for correlated nuclear
matter versus $k/k_F$ for various values of the correlation
parameter $k_Fc$ (b) The same,  but on a logarithmic scale, for
$n(x)$.} \label{}
\end{figure}
%%%%%%%%%%%%%%%%%%%%%%%%%%%%%%%%%%%%%%%%%%%%%%%%%%%%%%%%%%%%%%%%%%%%%%
%%%%%%%%%%%%%%%%%%%%%%%%%%%%%%%%%%%%%%%%%%%%%%%%%%%%%%%%%%%%%%%%%%%%%%
%FIGURE-2
\begin{figure}
\centering
\includegraphics[height=7.0cm,width=7.0cm]{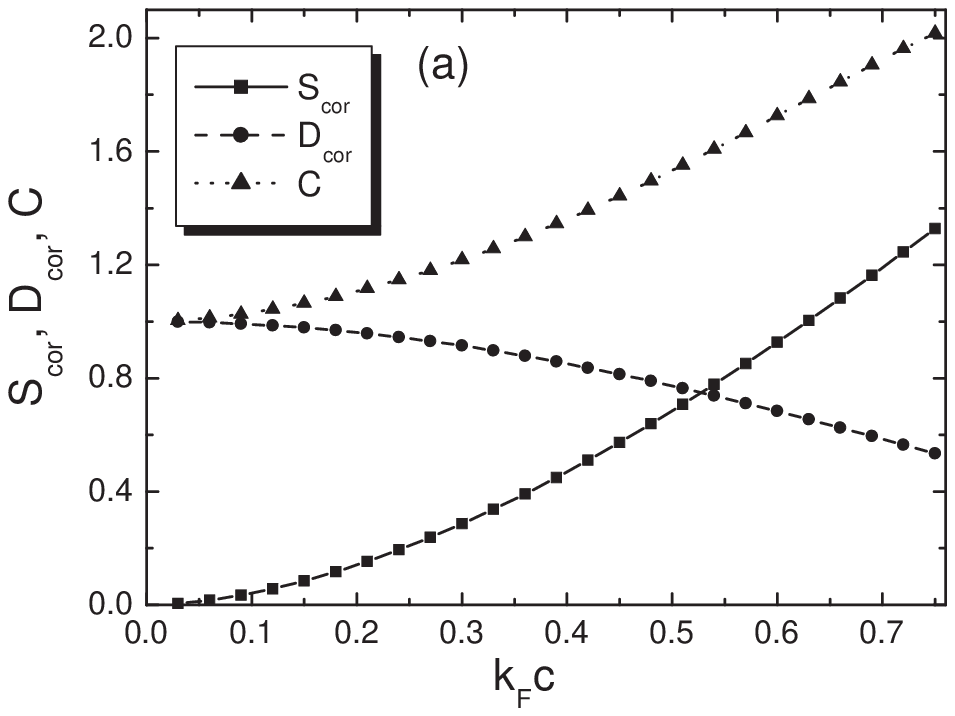}\
\includegraphics[height=7.0cm,width=7.0cm]{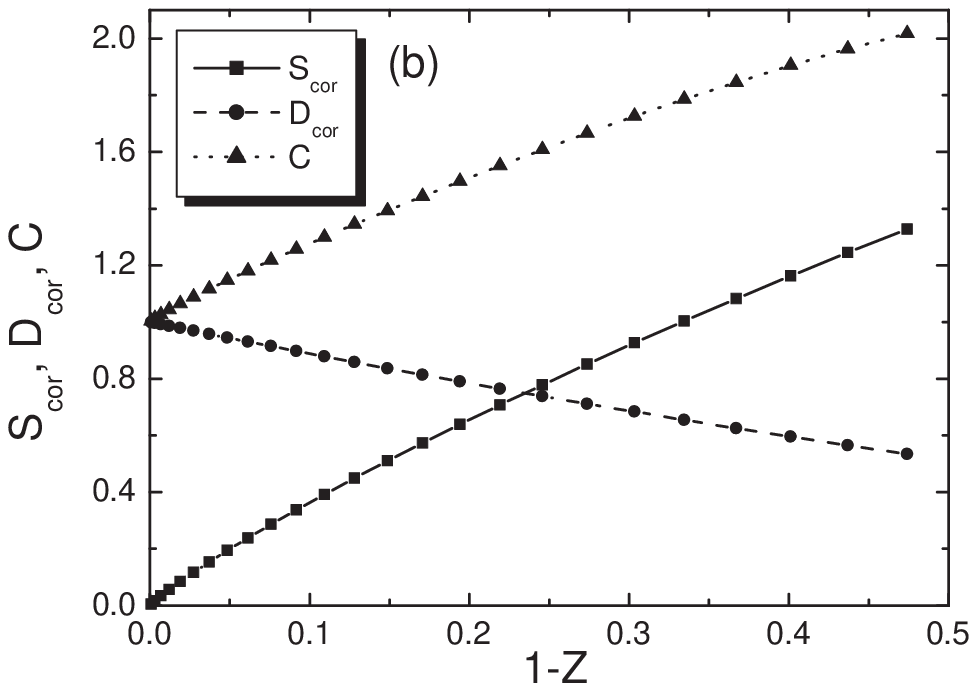}\
\caption{(a) $S_{cor}$, $D_{cor}$ and $C$ of nuclear matter versus
the the correlation parameter $k_Fc$ and (b) the discontinuity
parameter $(1-Z)$ } \label{}
\end{figure}
%%%%%%%%%%%%%%%%%%%%%%%%%%%%%%%%%%%%%%%%%%%%%%%%%%%%%%%%%%%%%%%%%%%%%%
%%%%%%%%%%%%%%%%%%%%%%%%%%%%%%%%%%%%%%%%%%%%%%%%%%%%%%%%%%%%%%%%%%%%%%
%FIGURE-3
\begin{figure}
\centering
\includegraphics[height=7.0cm,width=7.0cm]{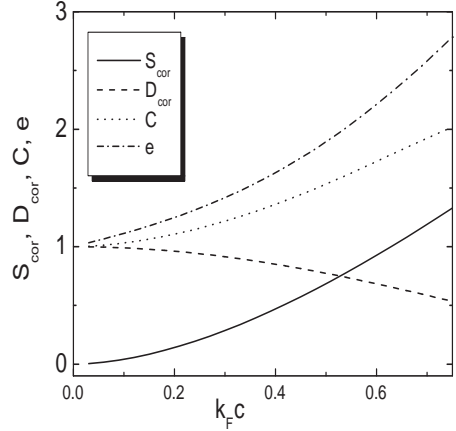}\
\caption{$S_{cor}$, $D_{cor}$, $C$ and the energy fraction $e$
(for definition see Eq.~(\ref{En-1})) versus the correlation
parameter $k_Fc$. } \label{}
\end{figure}
%%%%%%%%%%%%%%%%%%%%%%%%%%%%%%%%%%%%%%%%%%%%%%%%%%%%%%%%%%%%%%%%%%%%%%
%%%%%%%%%%%%%%%%%%%%%%%%%%%%%%%%%%%%%%%%%%%%%%%%%%%%%%%%%%%%%%%%%%%%%%
%FIGURE-4
\begin{figure}
\centering
\includegraphics[height=6.5cm,width=5.5cm]{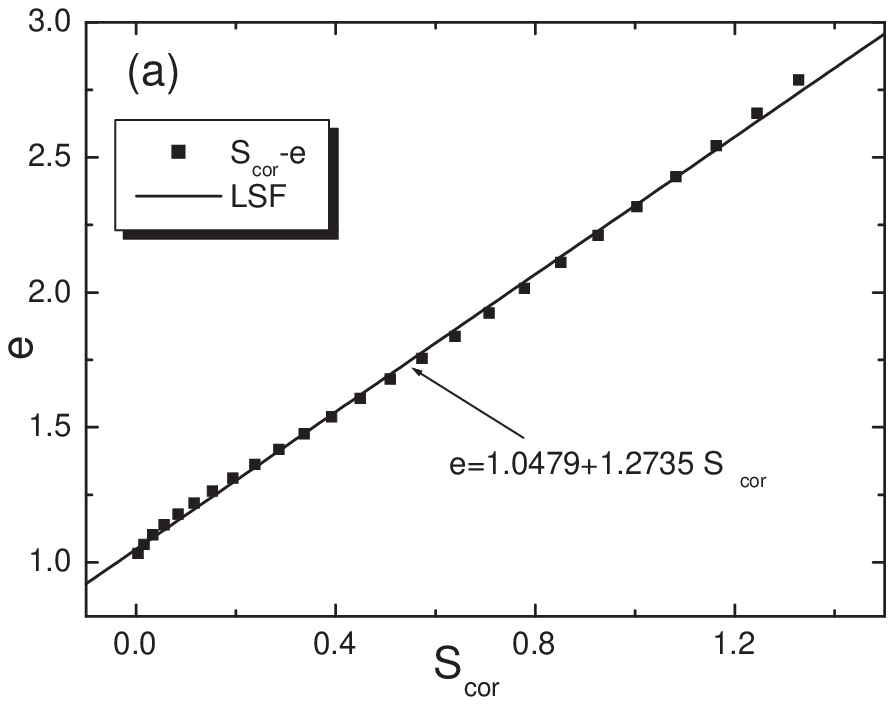}\
\includegraphics[height=6.5cm,width=5.5cm]{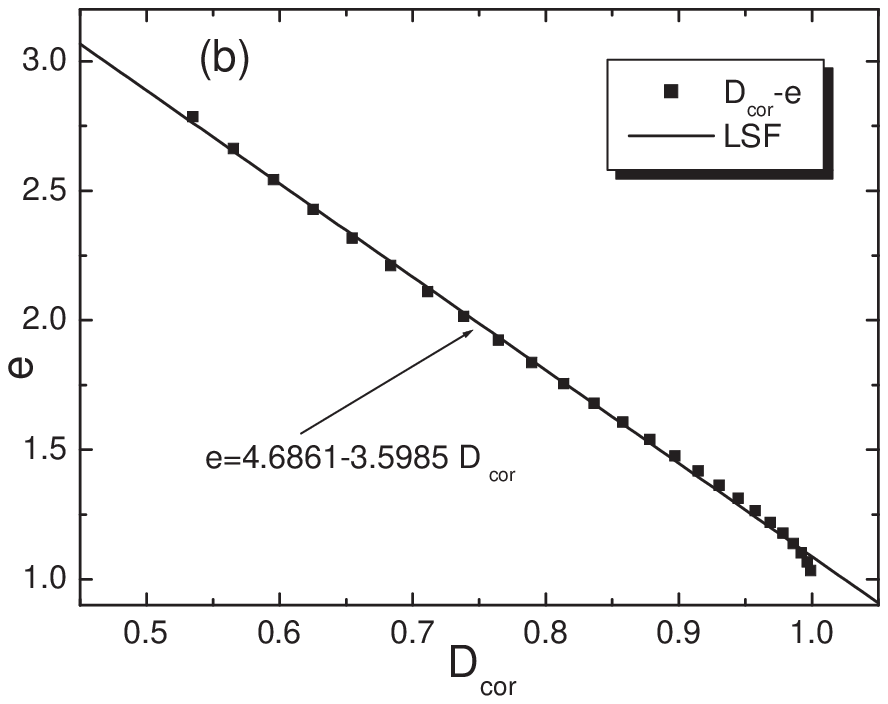}\
\includegraphics[height=6.5cm,width=5.5cm]{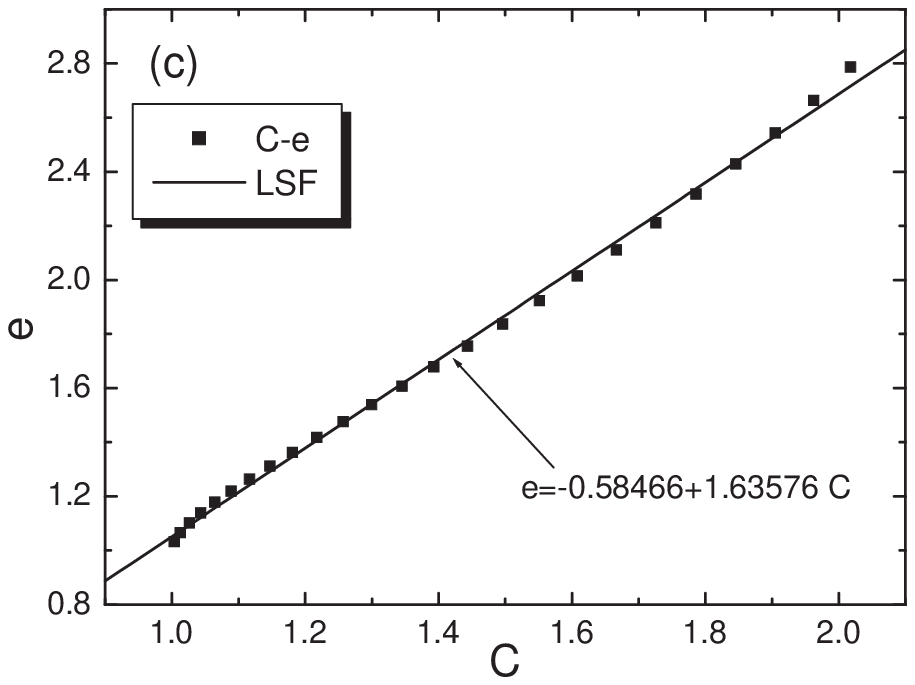}\
\caption{The energy fraction $e$ versus  (a) $S_{cor}$ (b)
$D_{cor}$ and (c) $C$. The lines corresponds to the expressions
(\ref{s-e}), (\ref{d-e}) and (\ref{c-e}), respectively   with the
parameters derived by the least squares fit method (LSF). }
\label{}
\end{figure}
%%%%%%%%%%%%%%%%%%%%%%%%%%%%%%%%%%%%%%%%%%%%%%%%%%%%%%%%%%%%%%%%%%%%%%

%%%%%%%%%%%%%%%%%%%%%%%%%%%%%%%%%%%%%%%%%%%%%%%%%%%%%%%%%%%%%%%%%%%%%%
%FIGURE-5
\begin{figure}
\centering
\includegraphics[height=7.0cm,width=7.0cm]{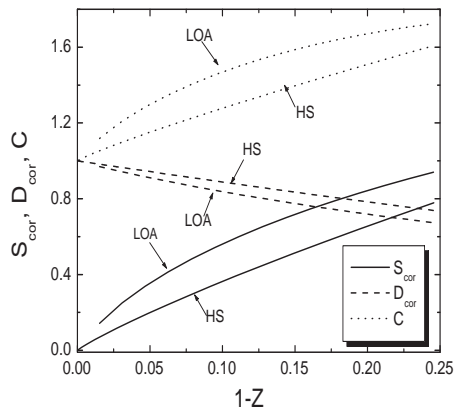}\
\caption{$S_{cor}$, $D_{cor}$ and $C$ of nuclear matter versus the
the discontinuity parameter $(1-Z)$  calculated by employing two
different methods: the Low Order Approximation (LOA) (see text)
and the hard sphere interaction (HS).} \label{}
\end{figure}
%%%%%%%%%%%%%%%%%%%%%%%%%%%%%%%%%%%%%%%%%%%%%%%%%%%%%%%%%%%%%%%%%%%%%%

\end{document}